\documentstyle[12pt]{article}
\catcode`@=11
\newcount\@tempcntc
\def\@citex[#1]#2{\if@filesw\immediate\write\@auxout{\string\citation{#2}}\fi
  \@tempcnta\z@\@tempcntb\m@ne\def\@citea{}\@cite{\@for\@citeb:=#2\do
    {\@ifundefined
       {b@\@citeb}{\@citeo\@tempcntb\m@ne\@citea\def\@citea{,}{\bf ?}\@warning
       {Citation `\@citeb' on page \thepage \space undefined}}%
    {\setbox\z@\hbox{\global\@tempcntc0\csname b@\@citeb\endcsname\relax}%
     \ifnum\@tempcntc=\z@ \@citeo\@tempcntb\m@ne
       \@citea\def\@citea{,}\hbox{\csname b@\@citeb\endcsname}%
     \else
      \advance\@tempcntb\@ne
      \ifnum\@tempcntb=\@tempcntc
      \else\advance\@tempcntb\m@ne\@citeo
      \@tempcnta\@tempcntc\@tempcntb\@tempcntc\fi\fi}}\@citeo}{#1}}
\def\@citeo{\ifnum\@tempcnta>\@tempcntb\else\@citea\def\@citea{,}%
  \ifnum\@tempcnta=\@tempcntb\the\@tempcnta\else
   {\advance\@tempcnta\@ne\ifnum\@tempcnta=\@tempcntb \else \def\@citea{--}\fi
    \advance\@tempcnta\m@ne\the\@tempcnta\@citea\the\@tempcntb}\fi\fi}
\catcode`@=12

\def\barr{\begin{array}}
\def\earr{\end{array}}
\def\beq{\begin{equation}}
\def\eeq{\end{equation}}
\def\bea{\begin{eqnarray}}
\def\eea{\end{eqnarray}}
\def\bmath{\begin{displaymath}}
\def\emath{\end{displaymath}}
\def\bq{\begin{quote}}
\def\eq{\end{quote}}
\def\nn{\nonumber\\}
\def\Re{\mbox{Re}}
\def\Im{\mbox{Im}}
\def\cA{{\cal A}}

\def\cL{{\cal L}}

\def\cO{{\cal O}}

\def\la{\lambda_a}
\def\lb{\lambda_b}
\def\lc{\lambda_c}
\def\PL{\mbox{P}_L}
\def\PR{\mbox{P}_R}

\def\slash#1{\setbox0=\hbox{$#1$}#1\hskip-\wd0\hbox to\wd0{\hss\sl/\/\hss}}

\def\gm{\mbox{g}}
\def\EE {\scriptscriptstyle E}
\def\veps{\varepsilon}
\def\g5{\gamma_5}
\def\lE{\lambda_{\EE }}
\def\lZ{\lambda_\ssz}
\def\Li2{\mbox{Li$_2$}}

\def\ssz{{\scriptscriptstyle Z}}
\def\ssw{{\scriptscriptstyle W}}
\def\sst{{\scriptscriptstyle T}}
\def\sse{{\scriptscriptstyle E}}
\def\roughly#1{\mathrel{\raise.3ex\hbox{$#1$
\kern-.75em\lower1ex\hbox{$\sim$}}}}
\def\lsim{\roughly<}

\def\mpla#1{{\em Mod.\ Phys.\ Lett.\ }{\bf A#1}}

\def\npb#1{{\em Nucl.\ Phys.\ }{\bf B#1}}
\def\plb#1{{\em Phys.\ Lett.\ }{\bf A#1}}
\def\plb#1{{\em Phys.\ Lett.\ }{\bf B#1}}
\def\prl#1{{\em Phys.\ Rev.\ Lett.\ }{\bf #1}}

\def\prd#1{{\em Phys.\ Rev.\ }{\bf D#1}}

\def\zpc#1{{\em Z.\ Phys.\ }{\bf C#1}}
\begin{document}

\begin{flushright}
NEIP-94-004\\[-0.2cm]
McGill-94/18\\[-0.2cm]
RAL/94-053\\[-0.2cm]
May 1994
\end{flushright}

\begin{center}
{\bf{\large ANOMALOUS VECTOR-BOSON COUPLINGS}}\\[0.3cm]
{\bf{\large IN MAJORANA NEUTRINO MODELS}}\\[2cm]
{\large C.P.~Burgess}$^{a}$
\footnote[1]{Permanent address: {\em Physics Department, McGill University,
3600 University St., Montr\'eal, Qu\'ebec, Canada, H3A 2T8.}}\footnotemark[2]
{\large and\hspace{0.13cm} A.~Pilaftsis}$^{b}$
\footnote[2]{E-mail addresses: cliff@physics.mcgill.ca,
pilaftsis@vax2.rutherford.ac.uk}\\[0.4cm]
$^{a}$ {\em Institut de Physique, Universit\'e de Neuch\^atel, 1 Rue
A.L.~Breguet,}\\
{\em CH-2000 Neuch\^atel, Switzerland.}\\[0.3cm]
$^{b}$ {\em Rutherford Appleton Laboratory, Chilton, Didcot, Oxon, OX11 0QX,
England.}
\end{center}
\vskip2cm
\centerline {\bf ABSTRACT}

We examine the contributions of
Majorana neutrinos to $CP$-violating $WWZ$ and $ZZZ$ self-couplings,
using a model in which sterile neutrinos couple to the
$W$ and $Z$ by mixing with a fourth-generation heavy lepton.
We find that the induced form factors can be as large as
$0.5\%$. The model satisfies all phenomenological bounds in
a natural way, including those due to the strong limits on the
neutron and electron electric dipole moments. Anomalous $CP$-odd
couplings of this size are unlikely to be observed at LEP200, but
might be detectable at NLC.

\newpage

\indent

The CERN Large Electron-Positron Collider (LEP), when operated at a
centre-of-mass energy of 200~GeV (LEP200), is expected to measure
the three-point gauge-boson self-couplings, and so to either
establish the non-Abelian structure predicted by the minimal
Standard Model (SM), or to observe deviations from this that might
signal the onset of new physics~\cite{GG,HPZH}. This prospect has
stimulated many detailed theoretical examinations of the prospects
for detecting these anomalous couplings at LEP200, or at a proposed
500~GeV linear $ee$
collider~(NLC)~\cite{GG,HPZH,EEWW,ABHIR,HMC,BBHR,HMM,CKP,DN,BFH}.
All of the explicit models that have been
proposed so far predict dauntingly small values ---  of order of
$10^{-3}$ in appropriate units --- for the $CP$-odd $WWZ$ and
$ZZZ$ form factors. Since measurements are expected only to be
sensitive to vector-boson self-couplings that are larger than
or of order $10^{-2}$ (NLC) or $10^{-1}$ (LEP200), this suggests
that any experimentally observed deviations from the
SM cannot be understood within the framework of a
perturbative, renormalizable field theory.
We believe it to be worthwhile to explore this conclusion
quantitatively, to see which scenarios maximize the expected
anomalous form factors. We focus here on $CP$-violating couplings,
since these are much easier to compute than are the $CP$-preserving
ones.\footnote[1]{See, however, refs. \cite{CPEVEN}.}

In this letter, we present a model which might be expected to produce
anomalous $CP$-odd couplings through new physics that is naturally
isolated from other observables, and so which is only quite weakly
constrained by current data. The predicted anomalous couplings
in this model therefore turn out to be comparatively large ---
up to $0.5\%$ --- although only of a size to be detectable at the NLC.
The model is based on supplementing the SM with a number of
electroweak-singlet sterile neutrinos, which then couple to the
electroweak bosons by mixing with a fourth-generation heavy neutrino
which is a member of a conventional weak isodoublet. This kind of
model has been previously considered as a potential
contributor~\cite{BS,Hansi} to the oblique parameters~\cite{STU}
of precision electroweak measurements,\footnote[2]{Although the
treatment of the oblique parameters in terms of the usual parameters
$S$, $T$ and $U$ is not justified for neutrinos with masses near the
electroweak scale~\cite{XYZ}, the bounds that were obtained in this way
are not expected to qualitatively change in a complete
treatment.} as well as a model for producing $CP$-odd~\cite{IKP}, and
other
quantum~\cite{Bernd} effects in the Higgs sector.

We first briefly describe the model. We require, in addition to the usual
SM particle content, a sequential heavy lepton, which we represent
with a weak isodoublet, and isosinglet
\bmath
L^0_L\ =\ \left( \barr{c} N^0_L\\ E^0_L \earr \right), \qquad E^0_R .
\emath
We imagine cancelling the electroweak anomalies of these fields by
including also a fourth generation of quarks, although these cannot
contribute to $CP$-violating anomalous gauge couplings at one loop,
and so play no role in what follows.
We finally add at least two right-handed sterile neutrinos,
which we collectively
denote as $N^0_{iR}, i=1,...n \ge 2$\footnote[3]{This type of model,
but with only one sterile neutrino, has been studied with the goal
of naturally accommodating a heavy fourth-generation neutrino~\cite{HP},
as well as for explaining the mass pattern of the light neutrinos~\cite{BM}.}.
At least two species of sterile neutrinos are required in order to permit
renormalizable $CP$-violating interactions amongst the neutrinos.

We assume for simplicity that the new sequential fourth-generation
particles mix only very feebly with the first three generations, as is
also required by global analyses of low-energy data~\cite{LL}. We do
not suppress, however, any mixing amongst the sterile neutrinos, or
between the sterile neutrinos and the fourth generation. The resulting
left-handed mass matrix, ${\bf M}$, for the heavy neutrinos then
takes the following form~\cite{IKP}:
\beq
{\bf M} \ =\ \ \left( \barr{cc} 0 & \mu^\sst \\ \mu & M \earr
\right),
\eeq 
where the first row and column correspond to the sequential
fourth-generation neutrino, $\nu_4 \equiv N^0_{0 L}$, which
we label in what follows with the subscript `0'. The rest of
the rows and columns represent
the various sterile neutrinos, $N^0_{iR}$. The quantities $\mu_i$
are generically complex numbers, but the matrix $M_{ij} =
M_i \delta_{ij}$ can without
loss be chosen to be diagonal, with real, nonnegative entries.
The mass eigenstates and eigenvalues are obtained by diagonalizing
${\bf M}$ by a unitary matrix ${\bf U}$ as follows:
\beq
{\bf U}^{\sst} {\bf M} {\bf U}\ \ =\ \ \hat{{\bf M}},
\eeq 
where the positive and diagonal matrix $\hat{{\bf M}}$
contains the mass eigenvalues of the heavy neutrinos $N_a,
\, a=0,...,n$, along its diagonal.\footnote[4]{We adopt here a
notation for which indices from the middle of the alphabet,
$i,j=1,...,n$, label the predominantly sterile neutrinos, while
indices from the beginning of the alphabet, $a,b=0,...,n$ also
include the dominantly isodoublet state, $N_0$. }
The spectrum of exotic fermions also includes the
heavy fourth-generation charged lepton, which we denote by $E$.
As long as these new particles are heavier than $M_\ssz/2$, so they
are not produced in $e^+e^-$ collisions at the $Z$ resonance,
their masses and couplings are largely unconstrained.

In terms of these mass eigenstates, the charged- and neutral-current
interaction of the heavy neutrinos become~\cite{ZPC}
\bea
\cL_{int}^\ssw &=& -\ \frac{g}{\sqrt{2}} \, W_\mu\
B_{\EE a}\
\bar{E}  {\gamma}^{\mu} \PL \ N_a \ + \
H.c. ,\\[0.3cm]
\cL_{int}^\ssz & =& -\ \frac{g}{4c_w} \, Z_\mu\
\bar{N}_a \gamma^\mu \Big[ C_{ab}\PL\ -\ C^\ast_{ab}\PR
\Big] N_b,
\eea 
where ${\PL}{}_{(R)}=(1-(+)\gamma_5)/2$, $c_w$ is the cosine of the
weak mixing angle, and the mixing matrices $B_{\EE a}$ and $C_{ab}$
are defined by
\beq
B_{\EE a}\ = e^{i\delta_E} U^{\ast}_{0a},\qquad
\quad
C_{ab}\ =\ U_{0a}U^{\ast}_{0b}.
\eeq 
In Eq.~(5), the phase $\delta_{\EE }$ is arbitrary and reflects
the freedom to rephase the charged lepton field $E$. This phase can
be used, for example, to ensure that $B_{\EE  1}$ is purely real.
The remaining quantities, $B_{\EE  i}$ for $i=2,...,n$, are then
generally complex, however, and their phases are the source of
$CP$-violation which we shall use.

For the purposes of illustrating the possible neutrino spectrum
and mixings, consider for a moment the case for which $|\mu_i| \ll M_j$,
for all $i$ and $j$. In this case $n$ of the mass-eigenstate neutrinos
are predominantly sterile, $N_i$ ($i=1,...,n$), and have masses $M_i +
O(\mu^2/M)$. The remaining neutrino, $N_0$, is lighter, having mass
$m_0$, where $m_0$ is the modulus of the following complex sum:
$\sum_i \mu^2_i / M_i \equiv m_0 \, e^{2i \delta}$. The phase of this
sum we call $2\delta$. With this
notation (and neglecting contributions of order $\mu^2/M^2$)
the mixing angles are:
\bmath
U_{00} = i e^{i \delta}, \qquad U_{0i} = i e^{i \delta} \, {\mu^\ast_i
\over M_i}, \quad (i=1,...,n).
\emath

In what follows we do not wish to make the assumption that the
$\mu_i$ are much smaller than the $M_i$. In this, the general case,
it is more fruitful to work directly with the neutrino masses and
mixings as our free parameters, keeping in mind that these are
restricted by the following general identities~\cite{ZPC,APetal}:
\beq
C_{ab} = B^\ast_{\EE a} B_{\EE b},\quad
\sum\limits_{c=0}^{n} m_c C_{ac} C_{bc} =  0,\quad
\sum\limits_{c=0}^{n} m_c B_{\EE c} C^\ast_{ca} =  0,\quad
\sum\limits_{c=0}^{n} m_c B_{\EE c}^2 =  0,
\eeq 
where $m_a$ denotes the mass of the $n+1$ Majorana neutrinos.

In order to be completely concrete, we specialize at this point
to the minimal case, for which we consider $n=2$ sterile neutrinos,
and so for which we have three heavy neutrino mass eigenstates.
In this case, as may be seen from Eqs. (6), the imaginary parts of
$B_{\EE 1}$ and $B_{\EE 2}$ are related to one other via
\beq
\Im\, B_{\EE 2}^2 \ \ = \ \ -\ \frac{m_1}{m_2}\; \Im\, B_{\EE 1}^2\, ,
\eeq 
where we have chosen the phase $\delta_{\EE}$ so that $\Im\, B_{\EE 0}=0$.

We now turn to the calculation of the $CP$-odd part of the
transition element $W^{-\nu}(p_1)\to Z^\mu(q)+ W^{-\kappa}(p_2)$
in this model. It is conventional to parametrize this in terms of
the following form factors~\cite{GG,HPZH}
\beq
\Gamma^{\mu\nu\kappa}_{\ssz\ssw\ssw}|_{CP-odd}\ =\ -f_\ssz(q^2) \,\veps^{\mu
\nu\kappa\rho} q_\rho - \frac{g_\ssz(q^2)}{M^2_\ssw} \, p^\kappa
\veps^{\mu\nu\sigma\rho} q_\sigma p_\rho + i h_\ssz (q^2) \, (q^\mu
{\gm}^{\nu \kappa} + q^\nu {\gm}^{\mu \kappa}),
\eeq 
where $p^2_1=p^2_2=M^2_\ssw$, $p=p_1+p_2$, and $f_\ssz$,
$g_\ssz$, $h_\ssz$ are $CP$-odd form factors. A similar analysis
for the matrix element $Z^\nu(p_1)\to Z^\mu(q)+ Z^\kappa(p_2)$
and assuming that the fields $Z^\nu$ and $Z^\kappa$ are on mass
shell gives~\cite{GG,HPZH}
\beq
\Gamma^{\mu\nu\kappa}_{\ssz\ssz\ssz}|_{CP-odd}\ =\
\frac{i \hat{h}_\ssz(q^2)}{M^2_\ssz} (q^\mu {\gm}^{\nu\kappa}
+ q^\nu{\gm}^{\mu\kappa}),
\eeq 
where $\hat{h}_\ssz$ is the anapole form factor for the $ZZZ$ vertex,
and ${\gm}_{\mu\nu}$ is the usual Minkowski-space metric.

For each of the form factors that appear in these expressions, there
is a similar one in which $Z^\mu(q)$ is replaced with a photon.
Of these, the two form factors, $f_\gamma$ and $g_\gamma$,
are particularly dangerous, contributing as they do to the neutron and
electron electric dipole moments (EDM's). As a consequence, these two are
experimentally constrained to be rather small~\cite{ABHIR,BFH}:
quantitatively they must satisfy~\cite{EDM} $f_\gamma (0)\ \lsim \
10^{-3}$ and $g_\gamma (0)\ \lsim\ 10^{-4}$. These bounds largely
preclude the possibility of observing $f_\gamma$ and $g_\gamma$
in $ee$ collisions for the forseeable future.

Any viable model for producing a sizable $CP$-violating anomalous
$WWZ$ or $ZZZ$ interactions, must therefore not also produce the
corresponding electromagnetic ones. One of the attractive
features of sterile-neutrino models is that this is ensured in
a completely natural way, because the $CP$-violating
$WW\gamma$ and $ZZ\gamma$ couplings {\it automatically} vanish
at one loop.
For the $WW\gamma$ vertex, this vanishing arises because
(in the absence of right-handed
charged currents~\cite{ABHIR,HMC}) any $CP$-violating phase in
the $W$--fermion coupling cancels between the two $W$ vertices.
Similar arguments hold for the $ZZ\gamma$ coupling. In this case
the vanishing of the one-loop $CP$-odd $ZZ\gamma$ form factors
is a consequence of the flavour-diagonal nature of the
the $Z$ and $\gamma$ couplings to the charged leptons and quarks,
as well as the absence of direct neutrino--photon couplings.
Other contributions to light fermion EDM's are precluded by the
assumed absence of mixing between the heavy and light leptons.

The same arguments do not rule out anomalous $WWZ$ and $ZZZ$ couplings
however. The difference is due to the possibility of having $CP$-violation
and neutrino flavour changes at the $Z$--fermion vertices. At one
loop only $f_\ssz$ and the anapole form factor, $h_\ssz$, turn out to be
generated by Fig.~1(a)~\cite{BFH}. For the model at hand, we find
\beq
f_\ssz(q^2)\ =\ -\frac{\alpha_w}{8\pi c^2_w} \sum_{ab} \Im C^2_{ab}\
I(q^2,\la,\lb, \lE),
\eeq 
where
\bea
I(q^2,\la,\lb,\lE)&=& \int_0^1\int_0^1 dx dy\, y^2(1-2x)\Bigg[
3\ln\cA^W(q^2,\la,\lb,\lE)\nn
&& \qquad\qquad -\ \frac{q^2}{4M^2_\ssw} \left(
\frac{1-y^2(1-2x)^2}{\cA^\ssw(q^2,\la,\lb,\lE)} \right)\Bigg],\\
\cA^\ssw(q^2,\la,\lb,\lE)&=&\lE(1-y)+\lb xy +\la (1-x)y - y(1-y)\nn
&& \qquad \qquad -\frac{q^2}{M^2_\ssw} y^2x(1-x)\ -\ i\veps\, ,
\eea 
and the kinematic variables $\la$ and $\lE$ are defined as
\beq
\la\ =\ \frac{m^2_a}{M^2_\ssw}, \qquad \lE\ =\
\frac{m^2_{\sse}}{M^2_\ssw}.
\eeq 

The summation over neutrino species in Eq. (10) may be simplified
by using the identities of Eq.~(6) to derive the following relations.
\bea
\Im \, C_{02}^2 &=& -\sqrt{{\lambda_1}/{\lambda_2}} \; \Im \, C_{01}^2,\nn
\Im \, C_{12}^2 &=& \sqrt{{\lambda_0}/{\lambda_2}} \;\Im \, C_{01}^2.
\eea 
These simplify Eq.~(10) to:
\bea
f_\ssz(q^2) &=& -\frac{\alpha_w}{4\pi c^2_w}\Im C_{01}^2 \; \Big[
I(q^2,{\lambda_0},{\lambda_1},\lE)\ -\ \sqrt{{\lambda_1}/{\lambda_2} }\;
I(q^2,{\lambda_0},{\lambda_2},\lE)\nn
&& \qquad \qquad +\ \sqrt{{\lambda_0}/{\lambda_2}}
\; I(q^2,{\lambda_1},{\lambda_2},\lE) \Big].
\eea 
Using $\Im\, C^2_{01} = \cO (1)$, in this expression gives the numerical
estimates of Tables 1 and 2 for LEP200 and NLC, respectively.
We find the largest values for $f_\ssz$ when the condition, $q^2 \simeq
(m_0 + m_1)^2$, for threshold effects is satisfied, and these
can be as large as $0.5\%$. For heavy neutrinos, {\em i.e.} $m_a
\gg M_\ssw$, we find smaller values: $f_\ssz \lsim 0.1\%$. Unfortunately,
LEP200 is likely to be unable to detect $CP$-violating anomalous $W$-
and $Z$-boson couplings that are smaller than $5-10\%$~\cite{CKP,DN},
and so these predictions are likely to be too small to be observed.
Nevertheless, $CP$-odd form factors as small as $0.5-1\%$ may be
accessible at NLC, given an upgrade in the luminosity or the adoption
of polarized $e^+$ and $e^-$ beams.

Our model also gives rise to an anapole form factor, $h_\ssz$, for the
coupling $WWZ$~\cite{BFH}, again from the graph of Fig.~1(a). We
find
\bea
h_\ssz(q^2) &=& \frac{\alpha_w}{4\pi c^2_w} \, \Im C_{01}^2 \;
\Big[ K(q^2,{\lambda_0},{\lambda_1},\lE) \ -\ \sqrt{{\lambda_1}/
{\lambda_2}} \; K(q^2,{\lambda_0},{\lambda_2},\lE) \nn
&& \qquad \qquad +\ \sqrt{{\lambda_0}/{\lambda_2}} \;
K(q^2,{\lambda_1},{\lambda_2},\lE) \Big],
\eea 
where
\bea
K(q^2,\la,\lb,\lE)&=& \int_0^1\int_0^1 dx dy\, y^2(1-2x) \Bigg[
\ln\cA^\ssw (q^2,\la,\lb,\lE) \nn
&& \qquad \qquad -\ \frac{q^2}{4M^2_\ssw}
\left(\frac{1-y^2(1-2x)^2}{\cA^\ssw(q^2,\la,\lb,\lE)} \right) \Bigg].
\eea 

Similarly, an anomalous anapole $ZZZ$ coupling, $\hat{h}_\ssz$, is induced
by the Feynman graph of Fig.~1(b)~\cite{BFH}:
\bea
\hat{h}_\ssz(q^2)\ &=&\ -\ \frac{\alpha_w}{8\pi c^2_w} \sum_{abc}
\Bigg[ \sqrt{\la\lb} \; \Im(C_{ac}C_{ab}C_{bc}^\ast) \;
L(q^2,\la,\lb,\lc)\nn
&& \qquad \qquad +\ \Im( C_{ac}^\ast C_{ab}C_{bc})
\; \hat{K}(q^2,\la,\lb,\lc) \Bigg],
\eea 
where
\bea
L(q^2,\la,\lb,\lc) &=& \int_0^1\int_0^1 dxdy \, \frac{y^2(1-2x)}
{\cA^\ssz(q^2,\la,\lb,\lc)},\\
\hat{K}(q^2,\la,\lb,\lc) &=& \int_0^1\int_0^1 dx dy \, y^2(1-2x)
\Bigg[ \ln \cA^\ssz (q^2,\la,\lb,\lc) \nn
&& \qquad -\ \frac{q^2}{4M^2_\ssw} \left(
\frac{1-y^2(1-2x)^2}{\cA^\ssz(q^2,\la,\lb,\lc)} \right) \Bigg], \\
\cA^\ssz(q^2,\la,\lb,\lc) &=& \lc(1-y)+ \lb xy + \la (1-x)y -
\lZ\; y(1-y) \nn
&& \qquad -\ \frac{q^2}{M^2_\ssw} y^2x(1-x)\ -\ i\veps\, ,
\eea 
and $\lZ = M^2_\ssz /M^2_\ssw$. Eq.~(18) can be significantly
simplified by judiciously using Eq.~(6). We find
\bea
\Im(C_{ac}C_{ab}C_{bc}^\ast) &=& C_{cc}\, \Im(C_{ab}^2),\\
\Im( C_{ac}^\ast C_{ab} C_{bc})&=& 0.
\eea 
Taking Eqs.~(14) and (22) into account, we arrive at our final
expression
\beq
\hat{h}_\ssz (q^2) \ =\ \frac{\alpha_w}{8\pi c_w^2}\,
\frac{\Im\, C^2_{01}}{\sqrt{\lambda_2}}\, \sum_{abcd} \veps_{abc}
\, C_{dd}\, \sqrt{\la\lb\lc}\; L(q^2,\la,\lb,\lambda_d),\quad
\eeq 
where $\veps_{abc}$ is the usual Levi-Civita tensor. When using this
expression to make numerical estimates, we assume that $C_{22} \ll 1$.
In this case Eq.~(24) simplifies to
\beq
\hat{h}_\ssz(q^2)\ \simeq\ \frac{\alpha_w}{4\pi c_w^2}\, \Im\,
C^2_{01} \, \sqrt{\lambda_0\lambda_1} \, \Big[\,
L(q^2,\lambda_0, \lambda_1, \lambda_0) \, -\,
L(q^2, \lambda_0, \lambda_2, \lambda_0)\, +\,
L(q^2,\lambda_1,\lambda_2,\lambda_0) \, \Big].\
\eeq 
We present our numerical results for the anapole form factors, $h_\ssz$
and $\hat{h}_\ssz$, at the relevant collider energies ($\sqrt{q^2}=200$~GeV
and 500~GeV) in Tables 1 and 2. As may be seen from the tables,
threshold effects can enhance the size of these couplings to the
level of $\sim 0.5\%$, which is on the edge of sensitivity at NLC.

Since the biggest contribution to the anomalous gauge couplings arises
due to threshold-mass effects of the Majorana neutrinos $N_0$, $N_1$, and
even these are at the edge of observability, one might expect
to pair produce the intermediate neutrinos via reactions such as
$e^+e^-\to N_a N_b$. Even if the heavy neutrinos should be sufficiently
long-lived to escape the detector --- such as if $m_0\lsim m_\sse$,
in which case $N_0$ cannot decay into charged leptons --- then it is likely
to be seen in measurements of the invisible $Z$ width at these energies.
This can be probed by looking for events in which a hard photon, radiated
from the initial electron/positron line, is seen to recoil against
something invisible. The rate for producing a light neutrino pair, such
as $N_0 N_0$, normalized by the total SM invisible width is
\beq
R_{mis.}\ =\ \frac{\sigma (e^+e^- \to N_0 N_0)}
{\sigma_{\scriptscriptstyle SM} (e^+e^-\to \hbox{invisible})} \
=\ |C_{00}|^2\, \beta_{N_0}\, \frac{3+\beta^2_{N_0}}{12},
\eeq 
where $\beta_{N_0}=(1-4m^2_{N_0}/q^2)^{1/2}$ is the velocity of the
outgoing $N_0$ in the centre of mass frame. For example, the rate for
producing a 50 GeV neutrino for $\sqrt{q^2}=200$~GeV would be $R_{mis.}\simeq
25\%$ if $C_{11}\simeq 1$. There is, however, a very narrow window
of masses for which $q^2$ is just on the lower rise of the threshold
enhancement, but for which there is insufficient energy for direct
neutrino production.

As can also be seen from the tables, the couplings are larger for smaller
values of the heavy charged-lepton mass (compare Table 1a with 1b, or
2a with 2b). If we restrict ourselves to the case where both the
charged lepton $E$, and the Majorana neutrino $N_0$, are too heavy to
be pair produced at the $q^2$ of interest, we are led to smaller results.
For example, we find in this case $f_\ssz \lsim 0.2\%$, yielding
$CP$-violating effects that are that much more difficult to detect.

In conclusion, we have demonstrated that Majorana-neutrino scenarios
based on the SM gauge group can predict an anomalous $WWZ$ coupling
$f_\ssz\ \lsim\ 0.5\%$. In principle, $CP$-violating effects due to
the dispersive (absorptive) parts of anomalous couplings can be observed
by looking at specific $CPT$-even ($CPT$-odd) observables in the decay
products of $W$-boson pairs~\cite{EEWW,CKP,DN}.
For example, effective $CPT$-even observables could be the
forward-backward asymmetry of the hardest jet when $W$ and $Z$ bosons decay
hadronically or $P$-odd momentum correlations between the initial electrons
and final charged leptons~\cite{CKP}.

\vskip1cm
\noindent
{\bf Acknowledgements.} A.P. wishes to thank the University of
Neuch\^atel, and C.B. would like to thank the Institute for
Theoretical Physics at Santa Barbara, for their kind hospitality.
This research was partially funded by N.S.E.R.C.\ of Canada,
les Fonds F.C.A.R.\ du Qu\'ebec, the Swiss National Foundation,
and the U.S. National Science Foundation (Grant No. PHY89-04035).

\newpage


\newpage

\centerline{\bf\Large Figure and Table Captions}
\vskip0.5cm

\newcounter{fig}
\begin{list}{\bf\rm Fig. \arabic{fig}: }{\usecounter{fig}
\labelwidth1.6cm \leftmargin2.5cm \labelsep0.4cm \itemsep0ex plus0.2ex }

\item Feynman graphs responsible for generating anomalous $CP$-violating
form factors in the vertices $WWZ$ and $ZZZ$.

\end{list}

\newcounter{tab}
\begin{list}{\bf\rm Tab. \arabic{tab}: }{\usecounter{tab}
\labelwidth1.6cm \leftmargin2.3cm \labelsep0.4cm \itemsep0ex plus0.2ex }

\item Numerical estimates of the anomalous vector-boson couplings
$f_\ssz$, $h_\ssz$, $\hat{h}_\ssz$ in units of $\Im C_{01}^2$ at
$\sqrt{q^2}=200$~GeV. We have used the values: {\bf (a)}
$m_0=m_E=50$~GeV and $m_2=1$~TeV, and {\bf (b)}
$m_0=m_E=100$~GeV and $m_2\gg 1$~TeV.

\item Numerical estimates of vector-boson $CP$-odd form factors
in units of $\Im C^2_{01}$ at $\sqrt{q^2}=500$~GeV.
We have assumed the values: {\bf (a)}
$m_0=m_E=50$~GeV and $m_2=1$~TeV, and {\bf (b)}
$m_0=m_E=300$~GeV and $m_2\gg 1$~TeV.

\end{list}

\newpage

\centerline{\bf\Large Table 1a}
\vspace{0.5cm}
\begin{tabular*}{13.91630391cm}{|r||rr|rr|rr|}
\hline
 & & & & & & \\
$m_1$ & $\Re f_\ssz$ & $\Im f_\ssz$ &  $\Re h_\ssz$ & $\Im h_\ssz$
& $\Re \hat{h}_\ssz$
& $\Im \hat{h}_\ssz$ \\
$[$GeV$]$ &  &  &  &  &  &  \\
\hline\hline
&&&&&& \\
100 & $2.3\ 10^{-3}$ & $2.5\ 10^{-3}$ & $4.2\ 10^{-4}$
    & $-1.6\ 10^{-3}$& $-6.7\ 10^{-4}$& $6.0\ 10^{-4}$\\
150 & $5.3\ 10^{-3}$ & 0              & $-3.9\ 10^{-3}$
    & 0              & $5.7\ 10^{-3}$ & 0             \\
200 & $2.1\ 10^{-3}$ & 0              & $-1.1\ 10^{-3}$
    & 0              & $1.3\ 10^{-3}$ & 0             \\
300 & $1.5\ 10^{-3}$ & 0              & $-6.9\ 10^{-4}$
    & 0              & $7.5\ 10^{-4}$ & 0             \\
400 & $1.2\ 10^{-3}$ & 0              & $-5.2\ 10^{-4}$
    & 0              & $5.6\ 10^{-4}$ & 0             \\
&&&&&&\\
\hline
\end{tabular*}

\vskip1cm

{\bf\Large Table 1b}
\hspace{1cm}
\vspace{0.5cm}
\begin{tabular*}{7.56254295cm}{|r||r|r|r|}
\hline
 & & & \\
$m_1$ & $\Re f_\ssz $ &  $\Re h_\ssz $ & $\Re \hat{h}_\ssz $ \\
$[$GeV$]$ &  &  &   \\
\hline\hline
&&& \\
200 & $9.2\ 10^{-4}$ & $-3.8\ 10^{-4}$& $3.7\ 10^{-4}$ \\
400 & $1.4\ 10^{-3}$ & $-5.1\ 10^{-4}$& $4.9\ 10^{-4}$ \\
600 & $1.5\ 10^{-3}$ & $-5.4\ 10^{-4}$& $4.8\ 10^{-4}$ \\
800 & $1.6\ 10^{-3}$ & $-5.6\ 10^{-4}$& $4.7\ 10^{-4}$ \\
1000& $1.7\ 10^{-3}$ & $-5.7\ 10^{-4}$& $4.4\ 10^{-4}$ \\
&&&\\
\hline
\end{tabular*}
\newpage

\centerline{\bf\Large Table 2a}
\vspace{0.5cm}
\begin{tabular*}{14.24352667cm}{|r||rr|rr|rr|}
\hline
 & & & & & & \\
$m_1$ & $\Re f_\ssz $ & $\Im f_\ssz $ &  $\Re h_\ssz $ & $\Im h_\ssz $
& $\Re \hat{h}_\ssz $
& $\Im \hat{h}_\ssz $ \\
$[$GeV$]$ &  &  &  &  &  &  \\
\hline\hline
&&&&&& \\
100 & $-1.1\ 10^{-3}$ & $6.2\ 10^{-4}$   & $8.4\ 10^{-3}$
    & $-3.3\ 10^{-4}$ & $-1.4\ 10^{-4}$  & $1.5\ 10^{-5}$   \\
200 & $-1.6\ 10^{-3}$ & $2.8\ 10^{-3}$   & $1.7\ 10^{-3}$
    & $-1.7\ 10^{-3}$ & $-5.3\ 10^{-4}$  & $1.9\ 10^{-4}$   \\
300 & $3.6\ 10^{-4}$  & $5.2\ 10^{-3}$   & $7.0\ 10^{-4}$
    & $-3.7\ 10^{-3}$ & $-7.3\ 10^{-4}$  & $8.3\ 10^{-4}$   \\
400 & $4.6\ 10^{-3}$  & $4.1\ 10^{-3}$   & $-3.0\ 10^{-3}$
    & $-3.5\ 10^{-3}$ & $8.8\ 10^{-4}$   & $1.2\ 10^{-3}$   \\
450 & $5.5\ 10^{-3}$  & 0                & $-4.2\ 10^{-3}$
    & 0               & $3.0\ 10^{-3}$   & 0               \\
500 & $2.9\ 10^{-3}$  & 0                & $-1.9\ 10^{-3}$
    & 0               & $1.2\ 10^{-3}$   & 0               \\
&&&&&&\\
\hline
\end{tabular*}

\vskip1cm

{\bf\Large Table 2b}
\hspace{1cm}
\vspace{0.5cm}
\begin{tabular*}{7.56254295cm}{|r||r|r|r|}
\hline
 & & & \\
$m_1$ & $\Re f_\ssz $ &  $\Re h_\ssz $ & $\Re \hat{h}_\ssz $ \\
$[$GeV$]$ &  &  &   \\
\hline\hline
&&& \\
400 & $3.6\ 10^{-4}$ & $-1.5\ 10^{-4}$& $1.4\ 10^{-4}$ \\
600 & $7.9\ 10^{-4}$ & $-3.0\ 10^{-4}$& $2.9\ 10^{-4}$ \\
800 & $1.1\ 10^{-3}$ & $-3.8\ 10^{-4}$& $3.7\ 10^{-4}$ \\
1000& $1.2\ 10^{-3}$ & $-4.3\ 10^{-4}$& $4.1\ 10^{-4}$ \\
1500& $1.5\ 10^{-3}$ & $-5.1\ 10^{-4}$& $4.5\ 10^{-4}$ \\
&&&\\
\hline
\end{tabular*}

\end{document}